\documentclass[aps,prl,twocolumn,amsmath]{revtex4-1}
\usepackage{bm}
\usepackage{mathrsfs}
\usepackage{graphicx}
\usepackage{dcolumn}

\DeclareMathOperator{\tr}{Tr}
\renewcommand{\vec}[1]{\mathbf{#1}}
\newcommand{\eqnref}[1]{Eq.~\eqref{#1}}

\begin{document} \title{Theory of NMR chemical shift in an electronic state
with arbitrary degeneracy}

\author{Willem Van den Heuvel}
\author{Alessandro Soncini}
\email{asoncini@unimelb.edu.au}
\affiliation{School of Chemistry, University of Melbourne, VIC 3010, Australia}

\begin{abstract}

We present a theory of nuclear magnetic resonance (NMR) shielding tensors for
electronic states with arbitrary degeneracy. The shieldings are here expressed in 
terms of generalized Zeeman ($g^{(k)}$) and hyperfine ($A^{(k)}$) tensors, of 
all ranks $k$ allowed by the size of degeneracy.  Contrary to recent proposals 
[T.~O. Pennanen and J. Vaara, Phys. Rev.  Lett. {\bf 100}, 133002 (2008)], our 
theory is valid in the strong spin-orbit coupling limit.  Ab initio calculations 
for the 4-fold degenerate $\Gamma_8$ ground state of lanthanide-doped fluorite 
crystals CaF$_2$:Ln (Ln = Pr$^{2+}$, Nd$^{3+}$, Sm$^{3+}$, and Dy$^{3+}$) show 
that previously neglected contributions can account for more than $50\%$ of 
the paramagnetic shift.

\end{abstract}
\maketitle

\textit{Introduction}.  Paramagnetic nuclear magnetic resonance (pNMR)
spectroscopy is a fundamental tool for probing static \textit{and} dynamic
\textit{local} magnetic properties of materials~\cite{CarrettaBook2007} and
metallo-proteins~\cite{Bertini}. pNMR plays a central role in the elucidation of 
the quantum dynamics of single-molecule
magnets and antiferromagnetic spin rings~\cite{NMRMolMag}.
Moreover, it is an increasingly central technique for
probing \textit{strong spin-orbit coupled}
electronic states in a dissipative environment, as shown by a
recent study of quantum tunneling processes~\cite{Ruben2009}, silent in
ac-susceptibility experiments, in lanthanide-based (Dy$^{3+}$ and Tb$^{3+}$)
molecular nanomagnets.

Despite the central role of pNMR in the development of new magnetic
materials, only very recently ab initio approaches have been developed for the
calculation of fundamental pNMR observables, such as nuclear shielding tensors~\cite{BookNMRandEPR2004}.
In particular, by generalizing the work of Moon and Patchkovskii~\cite{Moon2004}
on Kramers doublets, the paper by Pennanen and Vaara~\cite{Pennanen2008} 
stands out as the first comprehensive formulation 
of NMR shielding tensors in terms 
of molecular response properties, like the second-rank (first-rank in spin) EPR
$\mathbf{g}$ and hyperfine $\mathbf{A}$ tensors, routinely computed
via ab initio methods~\cite{Rinkevicius2003,BookNMRandEPR2004,
Soncini2007,Pennanen2008,Autschbach2011}. The work by Pennanen and
Vaara~\cite{Pennanen2008} aims at proposing an NMR-theory for arbitrary electronic 
ground state degeneracy. However, this task is only accomplished in~\cite{Pennanen2008} 
for the weak spin-orbit coupling limit, and to date \textit{no theory fully accounts for
truly general (spin and orbital) electronic degeneracy}, as that
arising in strong spin-orbit coupled lanthanide nanomagnets.  

In this Letter we propose a general theory for the ab initio calculation of
pNMR chemical shifts as function of molecular tensors of all ranks consistent
with the size of degeneracy, which applies to arbitrarily strong spin-orbit coupling (SOC) 
limit, and any degree of electronic degeneracy.

\textit{Theory}.
In the linear response regime, the electron-induced 
magnetic field $\vec{B}^I_\mathrm{ind}$ experienced by a nucleus $I$ in a
molecule immersed in a magnetic field $\vec{B}$, is proportional to the field
itself, so that $\vec{B}^I_\mathrm{ind} =-\vec{B}\cdot \bm{\sigma}^I$. This
leads to an electron-mediated interaction energy between external field and the
magnetic dipole moment $\bm{\mu}^I$ of nucleus $I$  which is bilinear in the
field and the nuclear dipole, given by  $W^{\vec{B},\bm{\mu}^I} =
-\vec{B}^I_\mathrm{ind}\cdot\bm{\mu}^I$. The proportionality tensor
$\bm{\sigma}^I$ is the NMR shielding tensor, its trace being 
related to
the chemical shift
of nucleus $I$.

In a system with a degenerate ground state, the energy
$W^{\vec{B},\bm{\mu}^I}$ must be averaged among the thermally accessible
states resulting from the splitting of the degeneracy under the
experimental conditions, leading to $\langle
W^{\vec{B},\bm{\mu}^I} \rangle = -\langle \vec{B}^I_\mathrm{ind}\rangle
\cdot\bm{\mu}^I = B_{\alpha}\langle \sigma^I_{\alpha\beta} \rangle
\mu^I_{\beta}$, where Greek indices correspond here to Cartesian tensor
components, and a sum is implied over repeated indices. 
Then the field-independent shielding tensor is defined as
\begin{equation} \langle \sigma^I_{\alpha\beta} \rangle =\left.
\frac{\partial^2 \langle W^{\vec{B},\bm{\mu}^I} \rangle }{\partial
B_{\alpha}\partial\mu^I_{\beta}} \right|_{\substack{\vec{B}\rightarrow 0\\
\bm{\mu}\rightarrow 0}}
=\left. -\frac{\partial \langle
B^I_{\mathrm{ind},\beta}\rangle}{\partial B_{\alpha}}
\right|_{\vec{B}\rightarrow 0}.  \label{secondderiv} \end{equation}
A strategy to evaluate~\eqnref{secondderiv} in transition
metal complexes has been first discussed in~\cite{Kurland1970}.
Here we reformulate this strategy in terms of projection operators.

We label $\mathrm{H}_0$ the unperturbed Hamiltonian in the absence
of the time-odd fields $\vec{B}$ and $\bm{\mu}^I$.  The well-known microscopic
operators~\cite{NoteRelativistic}, linear and bilinear in $\vec{B}$ and $\bm{\mu}^I$, describing the
interaction between electrons and fields are given by~\cite{Abragam1961}
(i) the Zeeman term $\mathscr{H}_z=-\bm{\mu}_e\cdot\vec{B}$ where
$\bm{\mu}_e$ is the electronic magnetic moment and (ii) the four Ramsey
terms, describing the interaction between the electrons and $\bm{\mu}^I$.  
The Ramsey terms can be further partitioned in a contribution
linear in $\bm{\mu}^I$, and a contribution that is bilinear in
$\vec{B}$ and $\bm{\mu}^I$. We collect the terms linear in
$\bm{\mu}^I$ in $\mathcal{F}_{\beta}^I\mu^I_{\beta}$, where the time-odd
operator $\mathcal{F}_{\beta}^I
=\mathcal{F}^I_{1,\beta}+\mathcal{F}^I_{2,\beta}+\mathcal{F}^I_{3,\beta}$ is
the sum of the nuclear-orbit ($\mathcal{F}^I_{1,\beta}$), Fermi-contact
($\mathcal{F}^I_{2,\beta}$) and spin-dipolar ($\mathcal{F}^I_{3,\beta}$)
terms~\cite{Abragam1961}. The bilinear term is 
$B_{\alpha}\,\mathcal{D}^I_{\alpha\beta}\,\mu^I_{\beta}$, where the
time-even operator $\mathcal{D}^I_{\alpha\beta}$ describes 
diamagnetic shielding in closed-shell molecules~\cite{Abragam1961,BookNMRandEPR2004}. 

The thermally averaged field at the position
of nucleus $I$ is obtained by averaging the total hyperfine field:
\begin{equation}\label{master}
\langle B^I_{\mathrm{ind},\beta} \rangle = 
 -\frac{\tr\bigl[\rho(\mathcal{F}_{\beta}^I
 +B_{\alpha}\mathcal{D}^I_{\alpha\beta})\bigr]}{\tr\rho},
\end{equation}
where $\rho=\exp[-\tilde{\beta}(\mathrm{H}_0+\mathscr{H}_z)]$ is the density
operator and $\tilde{\beta}=(k_\mathrm{B}T)^{-1}$. Next, we expand $\rho$ to
first order in $\mathscr{H}_z$ \cite{Schwinger1948, Kurland1970} and retain
those terms of \eqnref{master} that are linear in $B_\alpha$.
The result is expressed as a sum over the energy levels $n$ (with degeneracy
$\omega_n$) of $\mathrm{H}_0$:
\begin{multline}\label{vanvleck2}
\langle \sigma^I_{\alpha\beta} \rangle =\frac{1}{\mathcal{Z}_0}
 \sum_{n} e^{-\tilde{\beta}\varepsilon_n}\tr\biggl[
  \tilde{\beta}\, P_n \mu_{e,\alpha}P_n\mathcal{F}^I_{\beta}P_n
 +P_n\mathcal{D}^I_{\alpha\beta}P_n\\
    -P_n\mu_{e,\alpha}\frac{Q_n}{\varepsilon_n
   -\mathrm{H}_0}\mathcal{F}^I_{\beta}P_n-P_n\mathcal{F}^I_{\beta}\frac{Q_n}
   {\varepsilon_n -\mathrm{H}_0}\mu_{e,\alpha}P_n\biggr ]
\end{multline}
where $P_n=\sum_{i=1}^{\omega_n}|n\,i\rangle\langle n\,i|$ is the projector on
level $n$, $Q_n=1-P_n$, and
$\mathcal{Z}_0=\sum_n\omega_n e^{-\tilde{\beta}\varepsilon_n}$. 
This general expression for the shielding tensor formally resembles 
the Van Vleck equation for magnetic susceptibility~\cite{Gerloch1975}.

A common situation consists of a system with only one thermally occupied degenerate 
energy level. The shielding tensor of such system can be decomposed as
$\langle\sigma\rangle=\sigma^\mathrm{p}+\sigma^\mathrm{r}$,
where the index $I$ has been dropped to ease the notation. The 
temperature-independent term $\sigma^\mathrm{r}$ is universal to degenerate and
non-degenerate states alike, and is known as the Ramsey shielding in the latter
case.  The effect of degeneracy is to shift $\sigma^\mathrm{r}$ by a
temperature-dependent amount $\sigma^\mathrm{p}$, the {\it paramagnetic shift},
which will be the focus of this Letter, and which, from
\eqnref{vanvleck2}, can be expressed as ($P_0$ is the projector on the ground state):
\begin{equation}\label{sigma1}
\sigma_{\alpha\beta}^\mathrm{p}=
\frac{\tilde{\beta}}{\omega}\tr (P_0\mu_{e,\alpha}P_0\mathcal{F}_{\beta}P_0).
\end{equation}

In Ref.~\cite{Pennanen2008}, $\sigma^\mathrm{p}_{\alpha\beta}$ was
calculated from the parameters of an EPR spin Hamiltonian, expressed 
in the usual way as 
$\mu_\mathrm{B}\vec{S}\cdot\vec{g}\cdot\vec{B} + \vec{S}\cdot\vec{A}\cdot\vec{I}$.
However, this approach is only correct if $S\leq 1$, i.e., for 3-fold and 2-fold
electronic degeneracies, but it is only an approximation for higher spin.~\cite{Griffith1960}  
Moreover, while the approximation works for pure spin states,
definition of spin Hamiltonians in the strong SOC limit needs additional caution.

To describe the strong coupling case we have in fact to resort to the
concept of {\it fictitious spin}, and regard the degenerate manifold as a spin multiplet
$S$, so that $2S+1=\omega$. 
Once the $\omega$ wavefunctions of the ground manifold have been optimized
via ab initio methods, the $\omega\times\omega$ matrix representation 
$\bm{\mathsf{X}}$ of {\it any} operator $X$ in the basis of these wavefunctions can 
be reproduced by an effective operator in the spin space~\cite{Griffith1960},
sum over irreducible tensor operators $S_{q}^{(k)}$ of rank $k$, with 
spherical components $q = -k,\dots,k$ labelled by Latin indices:
\begin{equation}\label{spinexpansion}
\bm{\mathsf{X}}=\sum_{k=0}^{2S}\sum_{q=-k}^k(-1)^q X^{(k)}_q
\bm{\mathsf{S}}_{-q}^{(k)}. 
\end{equation}
Here the matrices $\bm{\mathsf{S}}_{-q}^{(k)}$, representation of $S_{-q}^{(k)}$
on the fictitious spin basis, form an orthogonal basis for the vector space of
all complex square matrices of dimension $2S+1$.  The coefficients $X^{(k)}_q$
can thus be determined from the ab initio matrices $\bm{\mathsf{X}}$ by 
orthogonal projection:
\begin{equation}\label{projection}
X^{(k)}_q = \tr\left({\bm{\mathsf{S}}_{q}^{(k)}} \bm{\mathsf{X}}\right)
 \frac{2k+1}{\langle S||S^{(k)}||S\rangle^2}
\end{equation}

Our first aim is to reformulate the Curie shielding tensor~\eqnref{sigma1} 
in terms of spin Hamiltonian parameters valid in the strong SOC limit, 
as these parameters can be routinely measured in EPR, 
and computed via accurate ab initio methods.  To this end, we apply~\eqnref{projection} to
spin-decompose the three components of the \textit{microscopic} Zeeman and
hyperfine ``fields'' appearing in~\eqnref{sigma1}, leading to three
sets of $X$-numbers for each field. A numerical example of how a projection is 
carried out using~\eqnref{spinexpansion} and~\eqnref{projection} is reported in the EPAPS~\cite{EPAPS}.

Following usual notation we name the resulting generalized EPR
``tensors'' collecting these sets of numbers $g_{q\alpha}^{(k)}$ (Zeeman)
and $A_{q\alpha}^{(k)}$ (hyperfine). For $k=1$,
we obtain the usual EPR $\vec{g}$-tensor and $\vec{A}$-tensor previously
considered~\cite{Moon2004,Pennanen2008}.  For a general degeneracy, and
arbitrary SOC strength,  we obtain: 
\begin{align}
P_0\mu_{e,\alpha}P_0 & = -\mu_\mathrm{B}\sum_{k q} (-1)^q S_{-q}^{(k)}\,
g_{q \alpha}^{(k)} \label{zeemanSH} \\
P_0\mathcal{F}^I_{\beta}P_0 & = \frac{1}{g_\mathrm{N}\mu_\mathrm{N}}\sum_{k q} (-1)^q
S_{-q}^{(k)}\, A_{q \beta}^{(k)}\label{hyperfineSH}
\end{align}
with $k\leq 2S$ and odd \cite{footnote}, and $g_\mathrm{N}$ the $g$-factor of the
nucleus in question.

Substituting these expressions in \eqnref{sigma1} yields 
\begin{equation}\label{sigma1gA}
 \sigma^\mathrm{p}_{\alpha\beta} = -\frac{\mu_\mathrm{B}}{
 g_\mathrm{N}\mu_\mathrm{N}}\frac{\tilde{\beta}}{2S+1} \sum_{kq}  g_{q\alpha}^{(k)}
A_{q\beta}^{(k)\ast} \frac{\langle S||S^{(k)}||S\rangle^2}{2k+1}.
\end{equation}
This expression generalizes all formulas previously proposed in the literature
to the strong spin-orbit coupling limit, and to arbitrary degeneracy,
thus representing one of the main results of this Letter.  Interestingly, 
only products of same-rank tensors enter~\eqnref{sigma1gA}. 

Next, we look for an expression for $\sigma^\mathrm{p}$ that is more general
than~\eqnref{sigma1gA}, accounting for a time-even perturbation
$\mathrm{H}_1$ that weakly splits the $2S+1$ degeneracy, which represents a
common situation (see Figure 1). In cubic symmetry $\mathrm{H}_1$ may 
describe the vibronic coupling of a $\Gamma_8$ electronic state to a Jahn-Teller
active mode. For a pure spin degeneracy ($S \ge 1$) \cite{Pennanen2008}, 
$\mathrm{H}_1$ leads to
the EPR zero-field splitting (ZFS) Hamiltonian.

The same arguments~\cite{Griffith1960} leading 
to~\eqnref{zeemanSH} and~\eqnref{hyperfineSH} can be used to project $\mathrm{H}_1$ 
on the ground-manifold, irrespective of its microscopic origin, and map it into 
a spin Hamiltonian. This leads to 
$P_0 \mathrm{H}_1 P_0 = \sum_{k,q}(-1)^q D^{(k)}_{q} S^{(k)}_{-q}$, 
where $k$ is an even integer~\cite{footnote} ($k\le2S)$, and the generalized ``ZFS tensors''
$D^{(k)}_{q}$ are evaluated via~\eqnref{projection}.
For instance, the projector $P_0$  on an ab initio $\Gamma_8$ 
ground manifold is $P_0 = \sum_{q} \left|q\rangle\langle q\right|$, where  
$q = \kappa, \lambda, \mu, \nu$~\cite{GriffithBook}. 
The rotational properties of a spin basis quantized along a $C_4$ axis ($Z$) 
readily lead to the mapping~\cite{GriffithBook}: 
$|\kappa\rangle \rightarrow |3/2\rangle$,
$|\lambda\rangle \rightarrow |1/2\rangle$, 
$|\mu\rangle \rightarrow |-1/2\rangle$,
$|\nu\rangle \rightarrow |-3/2\rangle$, and, for an axial distortion,
to the spin Hamiltonian 
$P_{0}\mathrm{H}_1P_{0}\equiv D (S_{Z}^2-5/4)$, where $D=\langle\kappa|\mathrm{H}_1|\kappa\rangle = 
\langle\nu|\mathrm{H}_1|\nu\rangle =
-\langle\lambda|\mathrm{H}_1|\lambda\rangle=
-\langle\mu|\mathrm{H}_1|\mu\rangle$.
The paramagnetic shift for a split manifold can now be obtained only using the
spin basis, via three simple steps.

First, we diagonalize the $\omega\times\omega$ spin Hamiltonian $P_{0}\mathrm{H}_1P_{0}$, 
and find the states $|\psi_{\lambda a} \rangle$  and split energies 
$\varepsilon_\lambda$, where  $\lambda=1,\dots,\omega^{\prime}$ ($\omega^{\prime}\le 2S+1$) 
labels the new low-lying degenerate manifolds, while $a$ spans a single manifold.  
Next, the eigenstates $|\psi_{\lambda a} \rangle$ are used to build new spin projectors
$\Theta_{\lambda} =\sum_{a}|\psi_{\lambda a}\rangle\langle\psi_{\lambda a}|$, 
each associated to a degenerate manifold $\lambda$.
For the axially split $\Gamma_8$ case, as illustrated in Figure 1, $D S_{Z}^2$ 
is diagonal, $\Gamma_8$ splits into 2 Kramers doublets (KD), and the new 
spin projectors can be written as ($D < 0$)
$\Theta_{\pm3/2}$ for the ground KD, and $\Theta_{\pm1/2}$ for the excited KD (see caption
to Figure 1).
\begin{figure}
\vspace{1cm}
\includegraphics[scale=0.52]{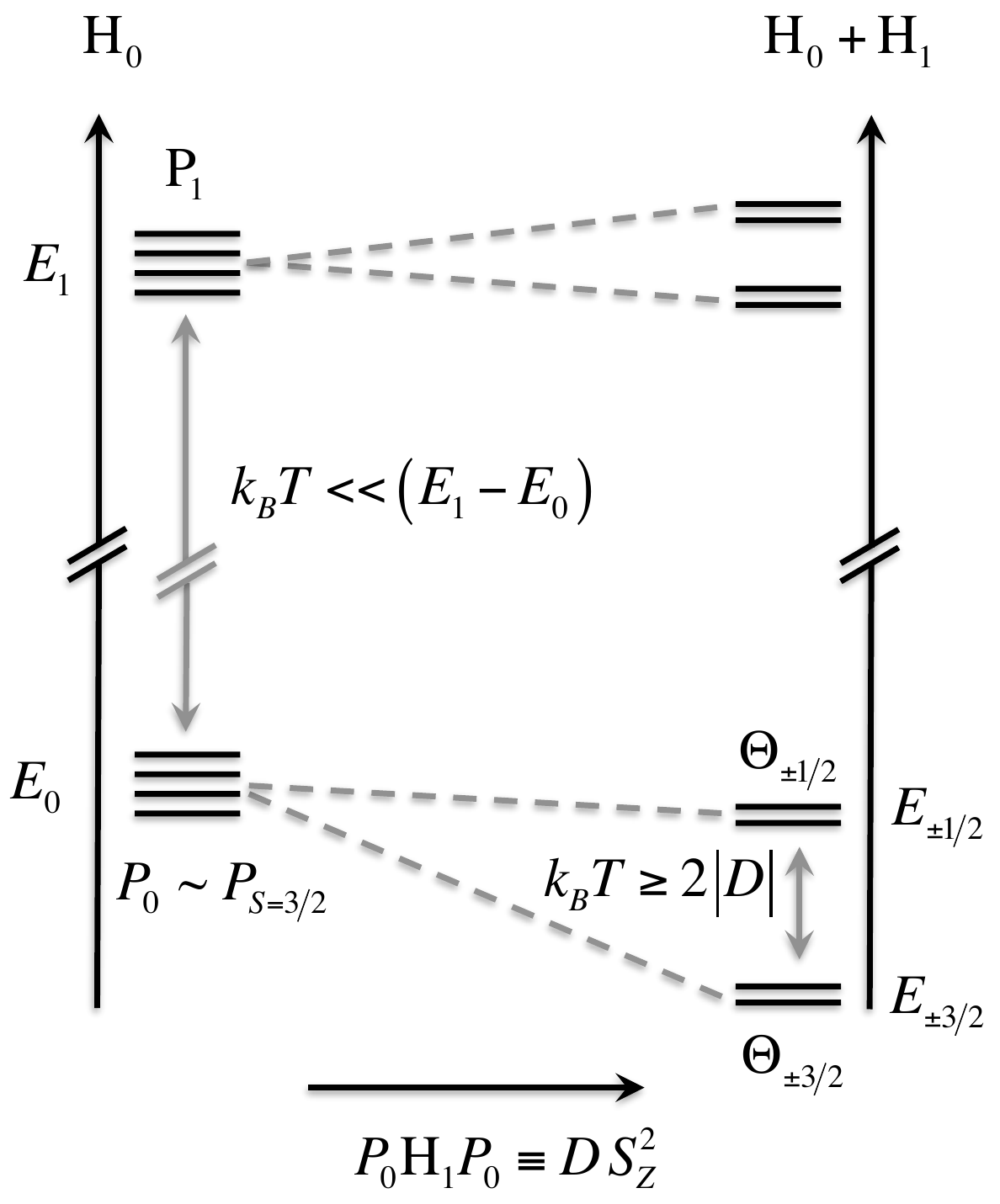}
\caption{Scheme of an axially split $\Gamma_8$ ground state and
associated projectors (see text). Left: $\Gamma_8$ ground state (GS)
of $\mathrm{H}_0$, with ab initio projector $P_0$ mapped to a spin ($S=3/2$) projector 
$P_{S=3/2}$.  Right: splitting induced by $\mathrm{H}_1$, described in GS by the spin
Hamiltonian $P_0\mathrm{H}_{1}P_0\equiv DS_Z^2$, whose eigenfunctions define the split projectors 
$\Theta_{\pm3/2} = |3/2\rangle\langle3/2|+|-3/2\rangle\langle-3/2|$,
$\Theta_{\pm1/2} = |1/2\rangle\langle1/2|+|-1/2\rangle\langle-1/2|$ .}
\end{figure}
Finally, the third step to obtain the paramagnetic shift for the split case is
to apply~\eqnref{vanvleck2}, but instead of microscopic Hamiltonians and projectors,
we now use the generalized Zeeman and hyperfine {\it spin Hamiltonians} 
($P_0 \mu_{e,\alpha}P_0$ and $P_0\mathcal{F}^I_{\beta}P_0$), 
and, crucially, the {\it spin projectors} defined on the split manifold ($\Theta_\lambda$). 

Following this strategy the paramagnetic shift of a split manifold
can be solely expressed as function of $g^{(k)}$ and $A^{(k)}$ tensors, 
the rank $k$ referring to the $(2S+1)$-manifold:
\begin{equation}\label{sigma1gAZFS}  \sigma^\mathrm{p}_{\alpha\beta}  =
\frac{\mu_\mathrm{B}}{g_\mathrm{N}\mu_\mathrm{N}}\frac{1}{\mathcal{Z}_0}
\sum_{kq,k^{\prime}q^{\prime}}(-1)^{q+q'}  g_{q\alpha}^{(k)}
\mathcal{Q}^{kk^{\prime}}_{qq^{\prime}}\left(T\right)
A_{q^{\prime}\beta}^{(k^{\prime})},  \end{equation}
where the partition function is defined over the
split manifold as $\mathcal{Z}_0 = \sum_{\lambda}^{\omega^{\prime}}
\omega_{\lambda} e^{-\tilde{\beta}\epsilon_{\lambda}}$, and the non-trivial
temperature dependent factor
$\mathcal{Q}^{kk^{\prime}}_{qq^{\prime}}\left(T\right) =
\mathcal{Q}^{(1)}_{kq,k^{\prime}q^{\prime}} +
\mathcal{Q}^{(2)}_{kq,k^{\prime}q^{\prime}}$:
\begin{align}
\mathcal{Q}^{(1)}_{kq,k^{\prime}q^{\prime}} &= -
\tilde{\beta}\sum_{\lambda=1}^{\omega^{\prime}} e^{-\tilde{\beta}
\epsilon_\lambda} \tr\left( \Theta_{\lambda} S^{(k)}_{-q} \Theta_{\lambda}
S^{(k^{\prime})}_{-q^{\prime}}\Theta_{\lambda} \right)\label{SSprojection1} \\
\mathcal{Q}^{(2)}_{kq,k^{\prime}q^{\prime}} &=
\sum_{\substack{\lambda=1\\ \mu\neq\lambda}}^{\omega'} e^{-\tilde{\beta} \epsilon_\lambda}
 \frac{2\Re\left[ \tr\left( \Theta_{\lambda} S^{(k)}_{-q}
\Theta_{\mu} S^{(k^{\prime})}_{-q^{\prime}}\Theta_{\lambda} \right)\right]}
{(\epsilon_{\lambda}-\epsilon_{\mu})} \label{SSprojection2} 
\end{align}
Note that \eqnref{sigma1gAZFS}
is different from the ZFS-expression
proposed in Ref.~\cite{Pennanen2008}, in that (i) it depends on
all possible products of tensors of all ranks allowed by the size of the
degenerate manifold (ii) the ``Curie term'',~\eqnref{SSprojection1}, cannot be
written in terms of averaged products of spin operators as
$\langle S_{q}^{(k)} S_{q^{\prime}}^{(k^{\prime})} \rangle$
(iii) \eqnref{sigma1gAZFS} contains a new ``orbital''
term, \eqnref{SSprojection2}, giving rise to a non-Curie temperature dependence.  
The new term is essential to achieve the correct
high-temperature limit, i.e.~\eqnref{sigma1gA}.
\eqnref{sigma1gAZFS}, generalization of~\eqnref{sigma1gA} to 
weak-split degeneracies, is the other main result of this Letter.

{\it Applications.}
We illustrate the theory by evaluating the NMR paramagnetic shift of strong
spin-orbit coupled dopant lanthanide nuclei in fluorite crystals, generally
denoted as CaF$_2$:Ln. Among the Ln-ions occupying a cubic
crystal site Pr$^{2+}$, Nd$^{3+}$, Sm$^{3+}$, and Dy$^{3+}$ are known to have a
4-fold degenerate ground state~\cite{Weber1964,Bernstein1979}, with symmetry $\Gamma_8$ in
the cubic double group $O^{*}$.  Our theory is essential to correctly describe NMR
for these nuclei.

Associating the $\Gamma_8$ states to a spin $S=3/2$ as seen previously,
from~\eqnref{zeemanSH} and~\eqnref{hyperfineSH} one obtains:
\begin{align}
P_0\mu_{e,\alpha}P_0&=-\mu_\mathrm{B}(g S_{\alpha} +g' W_{\alpha})
\label{ZeemanOct}\\
P_0\mathcal{F}^I_{\alpha}P_0&=\frac{1}{g_\mathrm{N}\mu_\mathrm{N}} (A S_{\alpha} 
+ A' W_{\alpha}),\label{hyperfineOct}
\end{align}
where
\begin{equation}
W_\alpha=5S^3_\alpha-\frac{41}{4}S_\alpha, \qquad \alpha=x,y,z
\end{equation}
are the components of $S^{(3)}$ that transform as a vector under
octahedral symmetry \cite{Abragam_EPR}.  Symmetry reduces the
number of non-zero tensor components to just four, denoted by $g$, $g'$, $A$,
and $A'$.  Next, by means of~\eqnref{sigma1gA}, the
paramagnetic shielding tensor (diagonal and isotropic in this symmetry) 
can be expressed as 
\begin{equation}\label{LNsigma}
\sigma^\mathrm{p}=
-\frac{\mu_\mathrm{B}}{g_\mathrm{N}\mu_\mathrm{N}}\tilde{\beta} \frac{15}{4}
\left(\frac{gA}{3} + 3\,g'A'\right)
\end{equation}

To calculate $\sigma^\mathrm{p}$ via~\eqnref{LNsigma}, we need 
the relevant rank-1 ($g$ and $A$) and rank-3 ($g'$ and $A'$) tensor 
components.  We choose the multiconfigurational wave function method CASSCF, 
followed by a non-perturbative calculation of the SOC 
as implemented in the code \textsc{molcas}~\cite{MOLCAS}.  The Ln-impurity and its first coordination
sphere (LnF$_8^{q-}$) are treated as an ab initio fragment embedded in the CaF$_2$ 
crystal potential, which is described by a combination of ab initio model potentials
(AIMP) \cite{seijo} and point charges in a finite cube centered on the Ln-ion
(see EPAPS~\cite{EPAPS} for further details).
\begin{figure}
\vspace{1cm}
\includegraphics[scale=0.85]{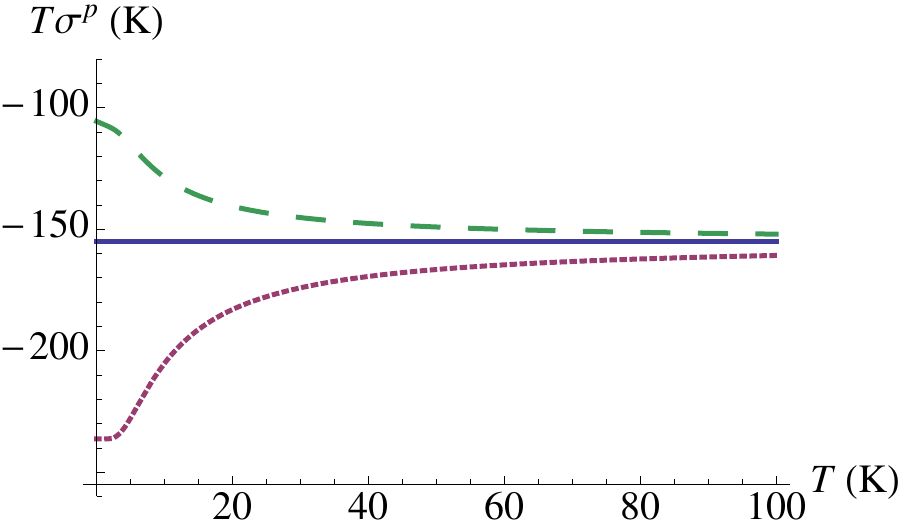}
\caption{(color online) Plot of $\sigma^p T$ \textit{vs.} $T$ for the Nd impurity nucleus in CaF$_2$:Nd$^{3+}$ 
crystal computed using the values in Table~\ref{table}. 
The solid (blue) line corresponds to $\sigma^p T$ for an exact $\Gamma_8$ 
ground state.  The dotted (red) and dashed (green) curves correspond to $\sigma^p_{\parallel} T$ and 
$\sigma^p_{\perp} T$, respectively, for an axially split ($D = -5\,$cm$^{-1}$) $\Gamma_8$ state,
calculated using~\eqnref{axialZFS}.}
\end{figure}

The ab initio matrix elements of the magnetic moment in the $\Gamma_8$
ground state were used to determine $g$~\cite{LiviuCeulemansBolvin2008} and $g'$.
Unfortunately, the calculation of the hyperfine constants is currently out
of reach of the CASSCF method for most systems.
Hence we chose to evaluate $A$ and $A'$ by assuming that $A/g =
A'/g'=A_J/g_J$, where $g_J$ and $A_J$ are the free-ion Land\'e factor and hyperfine
constant. This relation is exact in the limit of negligible $J$-mixing
\cite{Abragam_EPR_gA}, arguably a  good approximation for Ln-complexes.
Estimating $A_J/g_J$ from tabulated
experiments~\cite{Abragam_EPR_gA,Bernstein1979} allows us
to obtain $A$ and $A'$ from the computed values of $g$ and $g'$. All results
are in Table \ref{table}. The ab initio $g$ values for Pr$^{2+}$ and 
Dy$^{3+}$ compare reasonably well with available EPR data:
$g_{\mathrm{exp}}=-0.881$ and $g'_{\mathrm{exp}}=-0.173$ (Pr$^{2+}$)~\cite{Bernstein1979}, 
$g_{\mathrm{exp}}=4.4$ and $g'_{\mathrm{exp}}=0.19$ (Dy$^{3+}$)~\cite{Weber1964}.
Note that contributions arising from higher order EPR tensors ($g'$ and $A'$) 
in CaF$_2$:Nd$^{3+}$ (CaF$_2$:Pr$^{2+}$) account for $54\%$ ($46\%$) of the shielding tensor~\cite{NoteOnTiOctahedral}.

Finally, if the cubic environment of the impurity undergoes 
Jahn-Teller axial distortion along $z$, described by the 
ZFS fictitious spin Hamiltonian $D(S_z^2-5/4)$, application 
of~\eqnref{sigma1gAZFS}, leads to the following components of 
the anisotropic Ln-shielding tensor in the split-$\Gamma_8$ state:
\begin{eqnarray}\label{axialZFS}
 \sigma^\mathrm{p}_{\parallel}   &=&
-\frac{\mu_B}{g_\mathrm{N}\mu_\mathrm{N}}\frac{\tilde{\beta}}{\mathcal{Z}_0}
\left\{ \frac{9}{2}(g+g')(A+A') \right.\nonumber \\
&&\left. +\frac{e^{-2\tilde{\beta}\left|D\right|}}{2}(g - 9 g')(A - 9 A') \right\} \nonumber\\
 \sigma^\mathrm{p}_{\perp}   &=&  -\frac{\mu_B}{g_\mathrm{N}\mu_\mathrm{N}}
\frac{\tilde{\beta}}{\mathcal{Z}_0}
\left\{ \frac{225}{8} g'A' \right.\nonumber \\
&&\left. +\frac{e^{-2\tilde{\beta}\left|D\right|}}{8}(4 g + 9 g')(4 A + 9 A')  \right\}\nonumber\\
&& -\frac{\mu_B}{g_\mathrm{N}\mu_\mathrm{N}}\frac{\tanh \left( \tilde{\beta}\left|D\right| \right) }{16 \left|D\right|}\times\nonumber\\
&& \times 3\; (2 g - 3 g')(2 A - 3 A')
\end{eqnarray}
Note that in~\eqnref{axialZFS} products of tensors of different ranks appear, 
and the last term of $\sigma^\mathrm{p}_{\perp}$ is the new ``orbital'' contribution~\eqnref{SSprojection2}.  
Figure 2 shows the non-trivial temperature dependence of 
$ \sigma^\mathrm{p}_{\parallel} $ and $ \sigma^\mathrm{p}_{\perp} $ for the weakly split case, 
compared to the simple Curie term for a $\Gamma_8$ ground state,
in CaF$_2$:Nd$^{3+}$. This non-Curie behavior provides an experimental strategy to probe small interactions
between nanomagnet and environment, by monitoring the paramagnetic shift as function of temperature.
\begin{table}
\caption{Ab initio (CASSCF) EPR parameters and NMR paramagnetic shift in the $\Gamma_8$ ground 
state of CaF$_2$:Ln$^{n+}$, compared with available experimental data. 
The hyperfine constants (10$^{-2}$ cm$^{-1}$) are divided by the nuclear $g_\mathrm{N}$ 
factor to make them isotope-independent. 
Values of $T\sigma^\mathrm{P}$ (K) are obtained from \eqnref{LNsigma}.
In parenthesis we report the ab initio shift that would be obtained if $g'$ and $A'$
were neglected.\label{table}}
\begin{ruledtabular}
\begin{tabular}{lcccccc}
& \multicolumn{1}{c}{$g$} & \multicolumn{1}{c}{$g'$} & 
\multicolumn{1}{c}{$A/g_\mathrm{N}$} & \multicolumn{1}{c}{$A'/g_\mathrm{N}$} & 
\multicolumn{1}{c}{$T\sigma^\mathrm{p}$}& \multicolumn{1}{c}{$T\sigma^\mathrm{p}$(exp)} \\
\hline
Pr$^{2+}$& -0.799 & -0.245 & -2.17& -0.666 & -106 (-57) & -94\footnotemark[1] \\
Nd$^{3+}$& -0.802 & -0.293 & -2.66 & -0.971& -155 (-71) &  - \\
Sm$^{3+}$& -0.233 & -0.0632 & -2.77 & -0.754& -36 (-21)  &  - \\
Dy$^{3+}$& 4.257 & 0.198 & 6.05 & 0.281 & -867 (-850)  & -906\footnotemark[1] \\
\end{tabular}
\footnotetext[1]{Calculated from EPR data reported in Ref.~\cite{Weber1964,Bernstein1979}}
\end{ruledtabular}
\end{table}

In conclusion, we presented a general theory of NMR paramagnetic shifts, valid for 
arbitrarily degenerate electronic states. The shielding is a function 
of generalized EPR Zeeman and hyperfine tensors of all symmetry-allowed ranks.
The significance of the approach has been demonstrated by high-level ab initio calculations.
We believe the theory will be valuable to interpret pNMR data for lanthanide-based magnetic materials, and for 
the development of ab initio methods for pNMR.

\begin{acknowledgments} 
A.S. acknowledges support from the Selby Research Award, and the Early Career Researcher 
grant scheme from the University of Melbourne.
\end{acknowledgments}


\begin{thebibliography}{12}
\bibitem{NoteRelativistic} Use of relativistic operators, more
appropriate for heavy elements, would not affect the validity of this theory.
\bibitem{NoteOnTiOctahedral} Our theory can be relevant for d-electrons as well. Octahedral 
Ti(III) has a $\Gamma_8$ ground state, where $g=g'=0$ (see~\cite{GriffithBook}, p. 271), 
but mixing of the excited E$_\mathrm{g}$ configuration by SOC in first order leads to
$g=-g'=\frac{4}{5}\left(\frac{\zeta}{\Delta}\right)$,
where $\zeta$ is the SOC constant and $\Delta$ the crystal field splitting (i.e. $g=-g'\approx 10^{-2}$).
In this case our theory is clearly needed to estimate the correct pNMR shift.
\end{thebibliography}
\end{document}